\documentclass[cameraready]{Interspeech}

\title{FreeSonic: Training-Free Temporal-Aware Decoupled Attention for Precise Audio Editing}

\author[affiliation={1}]{Yuxuan}{Jiang}
\author[affiliation={2}]{Mingyang}{Han}
\author[affiliation={3}]{Yusheng}{Dai}
\author[affiliation={4}]{Andong}{Wang}
\author[affiliation={1}]{Tianhong}{Zhou}
\author[affiliation={5}]{Jiaxin}{Ye}
\author[affiliation={2}]{Dongxiao}{Wang}
\author[affiliation={2}]{Haoxiang}{Shi}
\author[affiliation={2}]{Boyu}{Li}
\author[affiliation={2}]{Jun}{Song}
\author[affiliation={2}]{Cheng}{Yu}
\author[affiliation={2}]{Bo}{Zheng}
\author[affiliation={1}]{Weibei}{Dou}
\renewcommand{\authorsep}{,\\}
\author[affiliation={1}, correspondingauthor]{Zehua}{Chen}
\author[affiliation={1}, correspondingauthor]{Jun}{Zhu}

\address{
    $^1$ Tsinghua University, China
    $^2$ Alibaba Group, China
    $^3$ Monash University, Australia \\
    $^4$ Renmin University of China, China
    $^5$ Fudan University, China
}

\email{
    jiangyux25@mails.tsinghua.edu.cn,
    \{zhc23thuml, dcszj\}@mail.tsinghua.edu.cn
}

\keywords{audio editing, training-free, rectified flow}

\usepackage{comment}
\usepackage{multirow}

\begin{document}

\maketitle

\begin{abstract}
Text-to-audio (TTA) generation has made significant strides, yet achieving precise and consistent audio editing remains a major challenge. However, existing methods struggle to balance temporal consistency with background preservation. In this paper, we propose FreeSonic, a training-free framework leveraging the state-of-the-art Rectified Flow-based TangoFlux model. FreeSonic utilizes an optimized inversion-reverse process and joint text-audio attention maps for precise target segment extraction. For content editing, a novel scheduled attention decoupling confines modifications to target regions while preserving original acoustic context. Furthermore, task-oriented noise injection enhances versatility for tasks such as audio removal and non-rigid replacement. Extensive experimental results demonstrate that FreeSonic achieves a superior balance by providing a high-fidelity and efficient solution for precise and consistent audio editing.
Project and demos: \href{https://free-sonic.github.io/}{https://free-sonic.github.io/}.
\end{abstract}

\section{Introduction}

Recent advances in text-to-audio (TTA) generation ~\cite{liu2023audioldm,ghosal2023text,huang2023make,evans2024fast} have achieved significant success, allowing the production of diverse and high-quality audio according to given text prompts. The field has witnessed a paradigm shift from U-Net to DiT architectures, from diffusion models to flow-based models, and from mel-spectrogram VAEs to waveform-based VAEs~\cite{hung2024tangoflux,li2025meanaudio,evans2024long,lee2024etta}. Specifically, flow-based models such as Flux~\cite{flux2024} construct a straight probability flow from noise to audio, enabling faster generation with fewer sampling steps and reduced training cost~\cite{lipman2022flow}. DiT models~\cite{peebles2023scalable}, mainly built on pure attention architectures, have demonstrated superior generation quality and scalability compared to U-Net models. Concurrently, waveform-based VAEs directly encode waveforms into 1D latents that are temporally aligned with the waveform, avoiding the information loss introduced by mel vocoders~\cite{li2025bridge,li2026audio}. However, there is still a large gap between our need and existing methods in terms of consistent and high-fidelity audio editing.

In the context of text-conditioned audio editing, existing methods~\cite{jia2025audioeditor,xu2024prompt,manor2024zero} have achieved impressive performance in content modification, semantic replacement, and style transfer. However, the additive nature of audio, where multiple sounds often overlap in time, makes it difficult to satisfy two key requirements: \textit{temporal consistency} and \textit{background preservation}. Temporal consistency requires that changes are limited to the target regions while unedited parts remain highly consistent to the original. Background preservation requires that non-edited background sounds stay intact even when the foreground events are modified. Since current methods struggle to decouple these overlapping sounds, modifying a specific part often leads to unintended changes in the entire audio.
While some training-based methods~\cite{wang2023audit} attempt to enhance control, they are often constrained by the necessity of constructing complex triplet datasets~\cite{ungersbock2025sao}, and frequently rely on specialized architectural designs~\cite{tao2025mmedit,gao2025rfm} or external auxiliary models~\cite{lan2025guiding}, resulting in high computational overhead and limited flexibility.

\begin{figure*}[t]
\centerline{\includegraphics[width=17.0cm]{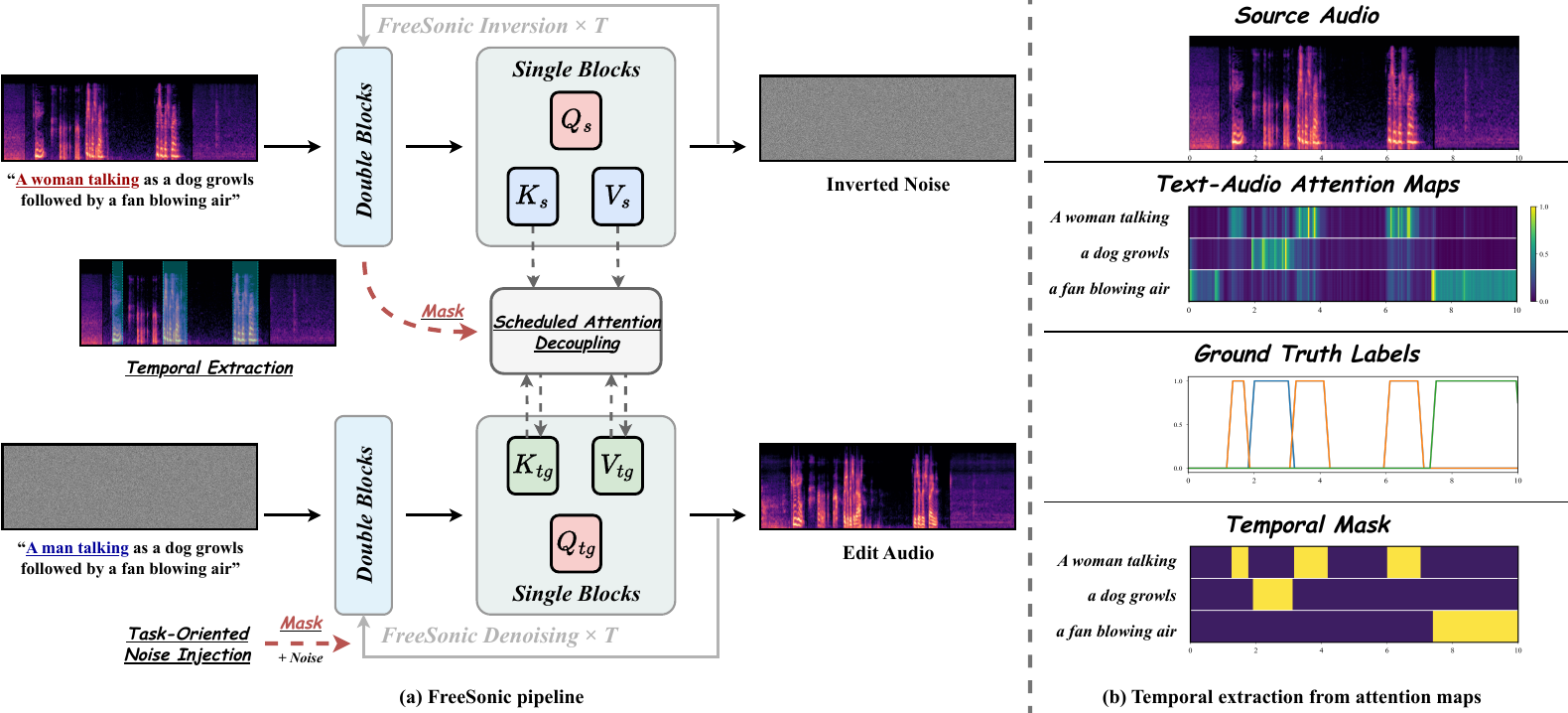}}
\caption{
Overview of the FreeSonic pipeline. (a) The editing workflow involves inversion and denoising, where Scheduled Attention Decoupling (Sec.~\ref{sec:sad}) and Task-Oriented Noise Injection (Sec.~\ref{sec:toni}) are applied for localized modification. (b) The temporal mask is extracted from text-audio attention maps (Sec.~\ref{sec:te}) during the first five inversion steps. By aggregating interaction scores in the double blocks, target segments are localized to guide editing while preserving the source background.
}
\label{fig:model}
\end{figure*}

In this paper, we propose FreeSonic, a training-free method designed to address the above challenges, enabling more consistent and precise audio editing, as shown in Figure~\ref{fig:model}. Our method is built upon TangoFlux~\cite{hung2024tangoflux}, a state-of-the-art TTA model that employs Rectified Flow (RF). To achieve reliable and consistent editing, FreeSonic focuses on the double blocks within the MM-DiT of TangoFlux. By leveraging the text-audio attention maps, we can accurately localize the segments corresponding to the edited textual prompts, ensuring that modifications are strictly confined to the intended regions while keeping the remaining content undisturbed. To perform precise editing, FreeSonic implements scheduled attention decoupling within the single blocks of MM-DiT. By leveraging KV features and the temporal mask, this strategy operates across different denoising stages to adaptively modulate the editing process in target regions. Specifically, it applies a scheduled fusion of source and target features to ensure precise content modification while fully injecting source KV features in non-edited areas to maintain maximum consistency. Furthermore, to enhance versatility across diverse editing tasks, we introduce task-oriented noise injection. Specifically, we inject random noise into the latent distribution of the target regions to reduce the influence of the original acoustic attributes. This strategy allows for more flexible modifications while maintaining structural and background consistency, enabling FreeSonic to handle challenging tasks such as audio removal and non-rigid replacement.

Crucial to the success of these editing mechanisms is the underlying stability provided by our optimized inversion-reverse process. Benefiting from the straight probability flow of RF models, this process solves the rectified flow ODE with reduced error, which enhances audio reconstruction quality and effectively preserves the structural information of the source audio. To the best of our knowledge, this represents the first attempt to introduce RF inversion into a training-free audio editing framework, providing a high-fidelity foundation for the subsequent evaluations.
Through extensive quantitative and subjective experiments, we demonstrate the effectiveness of FreeSonic across a wide range of audio editing tasks. Ablation studies confirm the contribution of each individual component, while efficiency analysis highlights its faster inference speed compared to existing models. By providing a flexible balance between temporal consistency and background preservation, FreeSonic demonstrates unique advantages in effectively addressing the intricate complexities of diverse real-world audio scenarios.

\section{Method}

\subsection{Preliminaries}

\subsubsection{Inversion-based Audio Editing}

Traditional inversion-based audio editing is primarily built upon diffusion models, such as DDIM and DDPM~\cite{jia2025audioeditor,xu2024prompt,manor2024zero,liang2025audiomorphix}. Given a source prompt $c$, the process maps the original audio latent into a noise representation via inversion, followed by reverse sampling guided by a target prompt $c^*$~\cite{dai2025latent,wang2025audiomog,dai2026omni2sound}. This paradigm aims to preserve the structural layout while incorporating semantic modifications. However, as the generative process is globally conditioned, altering the prompt shifts the entire latent trajectory, frequently leading to unintended deviations. Recently, inversion methods based on Rectified Flow have been proposed~\cite{wang2024taming,deng2024fireflow,xie2026dnaedit}, which benefit from straighter sampling trajectories and enhanced generation efficiency.

\subsubsection{TangoFlux Base Model}

FreeSonic builds on TangoFlux~\cite{hung2024tangoflux}, which employs Rectified Flow~\cite{liu2022flow} for TTA and is refined by CLAP-ranked preference optimization to enhance audio quality. Given a 1D audio latent $x_1$ and Gaussian noise $\tilde{x}_0 \sim \mathcal{N}(0, I)$, RF defines a linear probability path $x_t$ and the ground-truth velocity $v_t$ as:
\begin{equation}
x_t = (1 - t)x_1 + t\tilde{x}_0, \quad v_t = \frac{dx_t}{dt} = \tilde{x}_0 - x_1,
\end{equation}
where $t\in[0, 1]$. The model is trained as a velocity field $u(x_t, t; \theta)$ to predict $v_t$, conditioned on text embeddings $c_{text}$ from a Flan-T5~\cite{longpre2023flan} encoder and a duration encoding $c_{dur}$. The training objective minimizes the Flow Matching loss:
\begin{equation}
\mathcal{L}_{\text{FM}} = \mathbb{E}_{x_1, \tilde{x}_0, t} || u(x_t, t; \theta) - v_t ||^2 .
\end{equation}
This deterministic flow ensures a straight trajectory~\cite{esser2024scaling}, providing a stable foundation for training-free editing. 
During inference, noise $\tilde{x}_0$ is transformed back into the audio latent $x_1$.

\subsubsection{Attention Mechanism in MM-DiT-based TTA}
Unlike traditional U-Net architectures, TangoFlux employs the MM-DiT backbone consisting of multiple double and single blocks. In double blocks, text embeddings $c_{text}$ and audio latents $x_t$ are processed independently, whereas single blocks concatenate them into a unified sequence for joint modulation. Within these blocks, inter-modal interactions are captured via joint-attention. Specifically, the audio and text sequences are projected into queries $Q$, keys $K$, and values $V$ as:
\begin{equation}
Q = [Q_a; Q_t], \quad K = [K_a; K_t], \quad V = [V_a; V_t],
\end{equation}
where subscripts $a$ and $t$ denote the audio and text modalities, respectively. The attention output is then computed as:
\begin{equation}
\mathrm{Attention}(Q, K, V) = \text{softmax}\left(\frac{QK^T}{\sqrt{d}}\right)V.
\end{equation}
Due to the concatenation, each attention matrix can be decomposed into four components: audio-to-audio, text-to-text, text-to-audio, and audio-to-text attention. In TangoFlux, the text-to-audio and audio-to-text terms explicitly encode cross-modal alignment, enabling direct identification of temporal regions associated with specific semantic concepts. Meanwhile, the text-to-text and audio-to-audio components exhibit diagonal-dominant patterns, similar to observations in visual diffusion models~\cite{cai2025ditctrl}, indicating strong intra-modal temporal correlation essential for preserving coherent audio structure.

\subsection{Temporal-Aware Decoupled Attention}

\subsubsection{Text-Audio Attention for Temporal Extraction}
\label{sec:te}

To precisely localize editing regions, we exploit the natural alignment between the 1D latent structure of audio generative models and temporal progression~\cite{jiang2025freeaudio,wang2025audioatlas,jiang2025controlaudio}. While previous training-free editing approaches often rely on cross-attention manipulation to associate textual concepts with latent representations~\cite{jia2025audioeditor,xu2024prompt,manor2024zero}, the 1D structure of audio models offers a more direct mapping between latent positions and acoustic events. This property enables the precise localization of segments corresponding to textual descriptions by utilizing the joint-attention mechanism within the MM-DiT architecture, as illustrated in Fig.~\ref{fig:model} (b). Specifically, we analyze the text-audio attention maps that capture the interaction across all heads within the double blocks of the MM-DiT. By aggregating these scores, we derive a raw temporal importance map that is subsequently thresholded to produce a binary temporal mask $M \in \{0, 1\}^L$, where $L$ denotes the latent sequence length.

The mask $M$ is extracted during the first five steps of the inversion process~\cite{ouyang2025proedit}. At these initial stages, the latent representations maintain the strongest semantic correlation with the original audio and are least perturbed by noise, providing the most reliable alignment for localization~\cite{yin2025consistedit}. To address the coarse granularity of the downsampled latent space, we further refine $M$ through temporal dilation and smoothing. This refinement ensures comprehensive coverage of the target acoustic segment and eliminates internal discontinuities, facilitating seamless transitions and structural coherence during the editing process. The final extraction target is identified more accurately through specific prompt keywords or external masks.

\subsubsection{Scheduled Attention Decoupling}
\label{sec:sad}

To achieve stable and precise editing, we implement a three-stage scheduled decoupling strategy within MM-DiT single blocks. This approach progressively modulates the interactions between source and target KV features and the temporal mask $M$ across the early, intermediate, and late denoising stages.

\noindent \textbf{Stage 1: Feature Mixing.} In the early stage, the latent representation establishes the global layout and primary semantic attributes~\cite{jiang2025controlaudio,chen2025humo}. We regulate this process by decoupling and interpolating KV features within the self-attention mechanism. Let $(Q_{tg}^l, K_{tg}^l, V_{tg}^l)$ denote the query, key, and value at layer $l$ under the target prompt, and $(K_s^l, V_s^l)$ represent the reference features extracted from the source prompt. We mix source and target features to achieve semantic injection:
\begin{equation}
\begin{aligned}
\hat{K}_{tg}^l &= \delta K_{tg}^l + (1-\delta)K_s^l, \\
\hat{V}_{tg}^l &= \delta V_{tg}^l + (1-\delta)V_s^l,
\end{aligned}
\label{eq:mix}
\end{equation}
where $\delta$ is a scheduling coefficient that linearly transitions from $0.85$ to $1.0$ during this stage. Unlike fixed-weight methods~\cite{ouyang2025proedit}, this dynamic schedule ensures smoother acoustic transitions and more seamless fusion. To achieve precise region editing, we utilize a temporal mask $M$ to decouple the target from non-editing regions, applying a full injection of source features to the latter to ensure content consistency:
\begin{equation}
\begin{aligned}
\tilde{K}_{tg}^l &= M \odot \hat{K}_{tg}^l + (1-M)\odot K_s^l, \\
\tilde{V}_{tg}^l &= M \odot \hat{V}_{tg}^l + (1-M)\odot V_s^l.
\end{aligned}
\label{eq:mask}
\end{equation}
The updated latent representation is then computed as:
\begin{equation}
z^{l+1} = \mathrm{Attention}(Q_{tg}^l, \tilde{K}_{tg}^l, \tilde{V}_{tg}^l).
\end{equation}

\noindent \textbf{Stage 2: Temporal Control.} During the intermediate phase, we set $\delta$=1 in Eq.~\ref{eq:mix} to allow the target prompt to fully guide the semantic generation. The masking operation in Eq.~\ref{eq:mask} is strictly maintained to confine the editing process within the target region, ensuring the non-editing segments remain unchanged.

\noindent \textbf{Stage 3: Global Harmonization.} In the final steps, standard self-attention is restored to allow for global coordination across the entire sequence. By removing the KV constraints and the masking operation, the model can generate more harmonious transitions between the edited and unedited parts, ensuring the final audio as a whole is perceptually natural and coherent.

\subsubsection{Task-Oriented Noise Injection}
\label{sec:toni}

Deterministic inversion via RF effectively preserves the structural information of the source audio. However, this often leads to residual acoustic features in the latent space that interfere with modifications such as audio removal or non-rigid replacement~\cite{zhu2025kv}. To address this, we implement a task-oriented noise injection strategy that perturbs the latent exclusively within the temporal mask $M$ to break the deterministic dependency on the source content. Specifically, during the initial denoising steps $t \in [t_1, T]$, we update the latent $z_t$ as follows:
\begin{equation}
z'_t = (1 - M) \odot z_t + M \odot ((1 - \lambda_t) z_t + \lambda_t \eta),
\end{equation}
where $\eta \sim \mathcal{N}(0, I)$ and $\lambda_t$ is a linear scheduler. This scheduler scales the noise intensity from a predefined $\lambda$ at $t=T$ down to $0$ at step $t_1$, beyond which no further noise is injected. This unified noise injection strategy maintains plasticity in the edited regions while ensuring structural and background consistency in non-target areas, thereby enhancing editing performance in challenging scenarios like removal and non-rigid replacement.

\begin{table*}
\centering
\scriptsize
\caption{Quantitative evaluation of audio editing. Methods marked with $^*$ are training-based, while the others are training-free. \textbf{Bold} and \underline{underline} denote the best and second-best performance within each metric, respectively.}
\begin{tabular}{l|ccccc|ccccc|ccccc}
\toprule
\multirow{2}{*}{Method} 
& \multicolumn{5}{c|}{Add} 
& \multicolumn{5}{c|}{Remove} 
& \multicolumn{5}{c}{Replace} \\
\cmidrule(lr){2-6} \cmidrule(lr){7-11} \cmidrule(lr){12-16}
& FAD$\downarrow$ & KL$\downarrow$ & IS$\uparrow$ & FD$\downarrow$ & CLAP$\uparrow$ 
& FAD$\downarrow$ & KL$\downarrow$ & IS$\uparrow$ & FD$\downarrow$ & CLAP$\uparrow$ 
& FAD$\downarrow$ & KL$\downarrow$ & IS$\uparrow$ & FD$\downarrow$ & CLAP$\uparrow$ \\ 
\midrule
Ground Truth & - & - & - & - & 0.432 & - & - & - & - & 0.408 & - & - & - & - & 0.421 \\
\midrule
SDEdit       & 2.35 & 2.81 & 6.60 & 35.85 & 0.319 & 4.93 & 3.59 & 7.45 & 33.85 & 0.352 & 3.67 & 3.08 & 5.17 & 28.43 & 0.338 \\
AudioEditor  & 1.92 & 2.34 & 6.62 & 29.05 & 0.355 & 2.68 & 2.67 & \underline{8.14} & 31.83 & 0.395 & 2.29 & 2.54 & 6.28 & 24.98 & 0.362 \\
ZETA         & \underline{1.69} & 2.07 & \textbf{7.87} & 24.54 & \underline{0.368} & \underline{2.49} & 2.49 & \textbf{8.25} & 27.81 & 0.402 & \underline{2.27} & 2.34 & \textbf{7.02} & 28.62 & \underline{0.378} \\
SAO-Instruct* & 1.87 & \textbf{1.39} & 6.03 & \textbf{15.69} & 0.352 & 3.60 & \textbf{1.18} & 5.48 & \underline{19.09} & \underline{0.408} & 3.14 & \textbf{2.08} & 4.93 & \textbf{20.34} & 0.316 \\
FreeSonic    & \textbf{1.55} & \underline{1.58} & \underline{7.75} & \underline{19.59} & \textbf{0.374} & \textbf{1.95} & \underline{1.72} & 7.15 & \textbf{17.55} & \textbf{0.420} & \textbf{1.83} & \underline{2.21} & \underline{6.77} & \underline{22.97} & \textbf{0.424} \\
\bottomrule
\end{tabular}
\label{tab:obj}
\end{table*}

\begin{table*}
\centering
\scriptsize
\caption{Subjective evaluation of Audio Editing in terms of Quality, Relevance, and Faithfulness.}
\begin{tabular}{l|ccc|ccc|ccc}
\toprule
\multirow{2}{*}{Method} 
& \multicolumn{3}{c|}{Add} 
& \multicolumn{3}{c|}{Remove} 
& \multicolumn{3}{c}{Replace} \\
\cmidrule(lr){2-4} \cmidrule(lr){5-7} \cmidrule(lr){8-10}
& Quality$\uparrow$ & Relevance$\uparrow$ & Faithfulness$\uparrow$
& Quality$\uparrow$ & Relevance$\uparrow$ & Faithfulness$\uparrow$
& Quality$\uparrow$ & Relevance$\uparrow$ & Faithfulness$\uparrow$ \\
\midrule
Ground Truth  & 4.05 & 4.19 & 4.36 & 4.65 & 4.59 & 4.73 & 4.27 & 4.36 & 4.41 \\
\midrule
AudioEditor   & 3.34 & 3.28 & 3.14 & 3.35 & 3.26 & 3.31 & 3.28 & 3.15 & 3.26 \\
ZETA          & 3.53 & \underline{3.57} & 3.51 & \underline{3.61} & 3.51 & \underline{3.65} & \underline{3.62} & \underline{3.76} & 3.68 \\
SAO-Instruct* & \underline{3.70} & 3.51 & \textbf{3.68} & 3.55 & \textbf{3.73} & 3.59 & 3.58 & \textbf{3.81} & \underline{3.69} \\
FreeSonic     & \textbf{3.84} & \textbf{3.62} & \underline{3.59} & \textbf{3.71} & \underline{3.64} & \textbf{3.87} & \textbf{3.67} & 3.67 & \textbf{3.76} \\
\bottomrule
\end{tabular}
\label{tab:sub}
\vspace{-1em}
\end{table*}

\section{Experiments}

\subsection{Experimental Settings}

\subsubsection{Datasets and Baselines}
To evaluate audio editing performance, we construct a benchmark based on 10-second clips from the AudioCaps~\cite{kim2019audiocaps} test set and AudioSet Strong~\cite{hershey2021benefit}. We utilize annotations from the AudioCondition~\cite{guo2024audio} test set and additional high-quality audio-text pairs from FSD50K~\cite{fonseca2021fsd50k}, ESC-50~\cite{piczak2015esc}, and VGG-Sound~\cite{chen2020vggsound}, filtered via CLAP~\cite{wu2023large} to ensure semantic alignment. Our evaluation encompasses three primary tasks: addition (1,300 samples), removal (1,300 samples), and replacement (750 samples). For each sample, we manually craft a target caption based on the original content to guide the editing process, and provide corresponding natural-language instructions to facilitate the evaluation of instruction-tuned methods. We compare our method against training-free baselines including SDEdit~\cite{meng2021sdedit}, AudioEditor~\cite{jia2025audioeditor}, and ZETA~\cite{manor2024zero}, as well as the training-based SAO-Instruct~\cite{ungersbock2025sao}. For fair comparison, all methods are evaluated using their official open-source implementations, with SDEdit and ZETA built upon Stable Audio Open~\cite{evans2025stable}.

\subsubsection{Evaluation Metrics}

To quantitatively evaluate model performance, we employ several objective metrics, including Fréchet Audio Distance (FAD), Kullback–Leibler Divergence (KL), Fréchet Distance (FD), Inception Score (IS), and CLAP similarity~\cite{liu2023audioldm,wu2023large}, which together measure distribution alignment, generation quality, diversity, and audio–text semantic consistency. 
Efficiency is measured via the Real-Time Factor (RTF) on a single NVIDIA A800 GPU.
Additionally, we conduct subjective studies with 15 participants using Mean Opinion Score across three criteria: Quality, the perceptual excellence of the edited audio compared to the original; Relevance, how accurately the editing task was performed according to the instructions; and Faithfulness, the preservation of original acoustic components in the unedited regions.

\subsection{Quantitative Results}

We use the RF-Solver~\cite{wang2024taming} as a second-order ODE sampler with 25 denoising steps. The proposed three-stage scheduled attention decoupling is partitioned into 5 early, 5 intermediate, and 15 final steps. For task-specific noise injection, we set the intensity to 0.1, 0.4, and 0.25 for the Add, Remove, and Replace tasks, respectively, and set the noise injection cutoff step $t_1 = 5$. Table~\ref{tab:obj} summarizes the overall results. FreeSonic achieves superior results across most metrics, demonstrating robust capabilities in high-fidelity audio generation and semantic alignment. Our method outperforms the training-free baselines on most fidelity and semantic-alignment metrics. More importantly, FreeSonic surpasses the training-based SAO-Instruct* in several critical metrics, despite requiring no additional training. These results indicate that the proposed scheduled attention decoupling enables high-quality audio editing without the need for specialized fine-tuning or additional training data.

\subsection{Subjective Results}
As shown in Table~\ref{tab:sub}, FreeSonic demonstrates strong overall performance across the subjective metrics, confirming its perceptual effectiveness.
Additional qualitative samples across diverse scenarios are available on the project demo page.

\subsection{Ablation Study and Efficiency Analysis}

Table~\ref{tab:ablation} presents the ablation results for the key components of FreeSonic. The removal of the temporal mask, the substitution of scheduled decoupling with full KV replacement, or the exclusion of noise injection all result in a consistent degradation. These findings confirm that each component is essential for maintaining acoustic quality and ensuring semantic alignment with the target prompt.
Table~\ref{tab:efficiency} compares FreeSonic with other methods under both default and fixed NFE settings.
To ensure a fair evaluation, the NFE for training-free methods accounts for both inversion and denoising, where Classifier-Free Guidance (CFG) doubles the function evaluations per step.
We also evaluate the editing capability of first-order RF-Inversion~\cite{rout2024semantic} within our framework. The results indicate that our method attains better performance with lower computational overhead, highlighting its efficiency for high-quality audio editing.

\begin{table}[t]
\centering
\scriptsize
\caption{Ablation study on key components of FreeSonic.}
\begin{tabular}{l|ccc}
\toprule
Variant & FAD$\downarrow$ & KL$\downarrow$ & CLAP$\uparrow$ \\
\midrule
FreeSonic                               & 1.78 & 1.84 & 0.406 \\
\quad \textit{w/o temporal mask}        & 2.05 & 2.11 & 0.372 \\
\quad \textit{w full KV replacement}    & 1.96 & 2.07 & 0.383 \\
\quad \textit{w/o noise injection}      & 2.11 & 2.18 & 0.366 \\
\bottomrule
\end{tabular}
\label{tab:ablation}
\vspace{-0.5em}
\end{table}

\begin{table}[t]
\centering
\tiny
\caption{Efficiency analysis on audio editing. NFE includes inversion and CFG-based denoising to reflect latency.}
\setlength{\tabcolsep}{4pt}
\begin{tabular}{@{}l l l|cccccc@{}}
\toprule
Method & Model & Sampler & Steps & NFE$\downarrow$ & RTF$\downarrow$ & FAD$\downarrow$ & KL$\downarrow$ & CLAP$\uparrow$ \\
\midrule
AudioEditor  & Auffusion         & DDIM         & 100 & 300 & 2.744 & 2.30 & 2.52 & 0.371 \\
ZETA         & SAO               & DPM-Solver   & 100 & 300 & 1.711 & 2.15 & 2.30 & 0.383 \\
SAO-Instruct*& SAO               & DPM++ 3M SDE & 100 & 200 & 1.098 & 2.87 & 1.55 & 0.359 \\
\midrule
AudioEditor  & Auffusion         & DDIM         & 50  & 150  & 2.293 & 2.46 & 2.75 & 0.357 \\
ZETA         & SAO               & DPM-Solver   & 50  & 150  & 1.025 & 2.14 & 2.33 & 0.375 \\
SAO-Instruct*& SAO               & DPM++ 3M SDE & 75  & 150  & 0.720 & 2.85 & 1.53 & 0.342 \\
\midrule
FreeSonic    & TangoFlux         & RF-Inversion & 50  & 150 & 0.854 & 1.89 & 1.90 & 0.398 \\
FreeSonic    & TangoFlux         & RF-Solver    & 25  & 150 & 0.854 & 1.78 & 1.84 & 0.406 \\
\bottomrule
\end{tabular}
\label{tab:efficiency}
\vspace{-2em}
\end{table}

\section{Conclusion}
In this paper, we presented FreeSonic, a training-free audio editing framework built upon the Rectified Flow-based TangoFlux model. By leveraging joint text-audio attention maps for precise localization and implementing scheduled attention decoupling, FreeSonic ensures that modifications are strictly confined to target regions while preserving the original acoustic background. Task-oriented noise injection further enhances versatility across editing objectives.
Extensive quantitative and subjective experiments demonstrate that FreeSonic provides a training-free, high-performance and efficient solution for diverse audio editing tasks, achieving superior results in both acoustic fidelity and precise and consistent editing across various scenarios.

\section{Acknowledgments}
This work is supported by the Fundamental and Interdisciplinary Disciplines Breakthrough Plan of the Ministry of Education of China (No. JYB2025XDXM101), and the National Natural Science Foundation of China (62550004, U24A20342, U25B6003, 92570001).

\section{Generative AI Use Disclosure}
Generative AI tools were used for the sole purpose of editing and polishing the manuscript's language to improve clarity and readability. These tools were not used to generate any scientific content, data, or conclusions. All authors reviewed and edited the final manuscript and take full responsibility for the accuracy and integrity of the work.

\bibliographystyle{IEEEtran}
\bibliography{mybib}

\end{document}